Note

# DeepControl: 2D RF pulses facilitating $B_1^+$ inhomogeneity and $B_0$ off-resonance compensation in vivo at 7T

Mads Sloth Vinding[1], Christoph Stefan Aigner[2], Sebastian Schmitter[2,3], Torben Ellegaard Lund[1]

1, Center of Functionally Integrative Neuroscience (CFIN), Department of Clinical Medicine, Faculty of Health, Aarhus University, Denmark.
2, Physikalisch-Technische Bundesanstalt, Braunschweig and Berlin, Germany
3, University of Minnesota, Center for Magnetic Resonance Research, Minneapolis, MN, United States

*Corresponding author*:
Mads Sloth Vinding
E-mail: msv@cfin.au.dk
Address: Palle Juul-Jensen Boulevard 99, Bld. J117-154, 8200 Aarhus N, Denmark

**Abstract**

*Purpose:* Rapid 2D RF pulse design with subject specific $B_1^+$ inhomogeneity and $B_0$ off-resonance compensation at 7 T predicted from convolutional neural networks is presented.

*Methods:* The convolution neural network was trained on half a million single-channel transmit, 2D RF pulses optimized with an optimal control method using artificial 2D targets, $B_1^+$ and $B_0$ maps. Predicted pulses were tested in a phantom and in vivo at 7 T with measured $B_1^+$ and $B_0$ maps from a high-resolution GRE sequence.

*Results:* Pulse prediction by the trained convolutional neural network was done on the fly during the MR session in approximately 9 ms for multiple hand drawn ROIs and the measured $B_1^+$ and $B_0$ maps. Compensation of $B_1^+$ inhomogeneity and $B_0$ off-resonances has been confirmed in the phantom and in vivo experiments. The reconstructed image data agrees well with the simulations using the acquired $B_1^+$ and $B_0$ maps and the 2D RF pulse predicted by the convolutional neural networks is as good as the conventional RF pulse obtained by optimal control.

*Conclusion:* The proposed convolutional neural network based 2D RF pulse design method predicts 2D RF pulses with an excellent excitation pattern and compensated $B_1^+$ and $B_0$ variations at 7 T. The rapid 2D RF pulse prediction (9 ms) enables subject-specific high-quality 2D RF pulses without the need to run lengthy optimizations.

## 1. Introduction

Advanced radiofrequency (RF) pulses, e.g., 2D RF pulses designed with two spatial selective directions to select, e.g., a long cylindric or rectangular shape have as literature shows many applications, such as functional MRI (1,2), diffusion-weighted imaging (3), and 4D flow imaging (4).

RF pulse designers in general, navigate between matters of pulse types (e.g., excitation, refocusing or inversion); spatial (and/or spectral) target profiles (e.g., 2D RF, 3D RF, multi-band,...); inhomogeneous fields (single or parallel transmit $B_1$ ($B_1^+$) and $B_0$ maps); underlying field gradient waveforms (e.g., spirals, echo planar,...); system imperfections (5); constraints (e.g., power (6) and SAR (7)); and finally, which pulse design tool and physics model to adopt for the task, e.g., the Fourier-based small-tip-angle (STA) regime (8), a large-tip-angle (LTA) optimal control (OC) type (7,9), or a k-space (10) or spatial domain algorithm (11).

For 2D RF pulses, especially at ultrahigh field (UHF), it is important to accommodate a $B_0$ and a $B_1^+$ map in the pulse design to achieve a sharp profile and a homogenous flip angle distribution, respectively. Thus, tailoring the pulse to fit the patient (12). Universal, calibration free pulses are patient independent and may be a robust alternative, albeit with a trade-off in performance compared to tailored pulses (13). Nonetheless, it is essential for applicability of tailored pulses in clinics that the mapping stages and pulse computation times are fast (14).

To expedite the pulse computation time, several groups have recently investigated the potential of artificial intelligence (AI) in pulse design at 3 T. Tomi-Tricot et al. (15) introduced SmartPulse, which is a machine-learned calibration-free RF shimming approach. Shin et al. demonstrated deep reinforcement learning in root-flipping (16) for multiband Shinnar-LeRoux pulses, DeepRF$_{SLR}$ (17), and recently a self-learning machine following adiabatic RF design criteria, DeepRF (18). We recently showed rapid 2D RF pulse design for reduced-FOV imaging at 3 T by means of deep learning (DL) a fully connected neural network (NN) (14). There, we compared both STA and LTA pulse designs with DL-predicted pulses and found similar trends between predicted and (conventionally) computed pulse, and we showed preliminary data for $B_0$ inhomogeneity compensation.

In this study, we extend the method to include $B_1^+$ and $B_0$ inhomogeneity compensation in the brain at 7 T through a convolutional neural network. To the best of our knowledge these are the first in vivo, human experiments with AI-powered pulse design at 7 T.

## 2. Materials and methods

The principle of this method is described in more detail in (14). In brief, a vast training library is established, where each entry mimics a realistic pulse design situation: the 2D target pattern is defined, and artificial maps of $B_0$ off-resonances and single-channel $B_1^+$ in the image plane are defined with dynamic ranges corresponding to typical conditions. With supervised learning, a target method (TM) is then used to optimize 2D RF pulses for each library entry that achieve the target excitation profile under the given $B_0$ and $B_1^+$ circumstances. The target pattern and associated $B_0$ and $B_1^+$ maps are referred to as TM or DL *input*. The TM-computed single-channel 2D RF pulses are referred to as TM *output*, or DL *target*, i.e., what the AI framework is aimed to achieve, based on the DL input. The neural network is constructed such that it takes the DL input and returns a DL *output,* which is the 2D RF pulse it can achieve, based on the DL input. The training of the neural network is a minimization of the prediction root mean square error (RMSE) between the DL output and the DL target. Numerical operations were performed in MATLAB (Mathworks, Natick, MA, USA) on a 28-CPU core (Intel Xeon Gold 5120, 2.2 GHz) machine with 384 GB RAM and a Tesla P100 GPU (Nvidia Corp., Santa Clara, CA, USA).

### 2.1 Training library

The training library was set up to match reported $B_0$ offset conditions commonly observed at 7 T: the expected maximum $B_0$ offset frequency range is between ±50 and ±600 Hz, for example around 500 Hz has been observed in regions above nasal cavities at 7 T (19). One TM/DL input was a 64x64x3 matrix, where the 64x64 grid would intentionally span a 25-cm square (see Sec 2.2). The first 64x64 layer was the target excitation pattern, made from ten randomly placed points forming a polygon then closed morphologically. The script assured there would only remain one arbitrary shape with no holes, and it was assigned a target flip angle (FA) of 30°. The second layer constituted a $B_0$ map.

We used images randomly selected from the ImageNet database (20). Cut to 64x64 and gray-scaled, we shifted the pixel values by subtracting the median pixel value. From a $B_0$ map perspective, this induced a negative and positive offset range. After normalization, the pixels were then scaled by an arbitrarily selected maximum frequency offset in the range -600 to -50 Hz and +50 to +600 Hz.

The third layer was the (magnitude) $B_1^+$ map, again derived from the ImageNet database. This image was gray-scaled and normalized and would represent a $B_1^+$ map with magnitude sensitivities between 0 and 1.

Each layer in the 64x64x3 input matrices would be masked by a super ellipse shape, to mimic an object boundary, resulting effectively in 2919 spatial controls points. Our library consisted of 500k cases. As described in more detail in (14), we split the library into fractions: 60% for training, 20% for validation and 20% for final testing.

**2.2 Target method**

The gradient waveform accommodating the spatial selectivity of the 2D RF pulses formed an inward 16-turn spiral in the excitation k-space (21): duration 7 ms, dwell time $\Delta t = 10$ µs, peak gradient amplitude limit 40 mT/m, gradient slew rate limit 180 T/m/s, and field of excitation 25 cm. Hence, the spatial raster grid, defined by the training library input matrices of size 64x64 in the 2D plane, had a resolution of 3.9 mm.

The TM in this study was a fast, OC quasi-Newton procedure described in details in (7,22). RF pulses were constrained to 1 kHz (real/imaginary) amplitude. In order to produce 500k pulses in reasonable time, we exploited a modified, first-order approximate optimization gradient described in a forthcoming paper. We limited the number of available iterations to 50 and set the functional and step-size convergence tolerances to $1 \cdot 10^{-6}$.

We compared the normalized root-mean-square errors (NRMSE) of the resulting magnetizations obtained from Bloch simulations. As described in (14), this is a different measure of performance than that taking place during DL (see Sec 2.3).

**2.3 Neural network**

It was shown in (14) that three fully connected layers with rectified linear unit layers between could transfer a 2D target pattern from the input layer to a final regression layer, which resembled a 2D RF pulse similar to that of the TM. We also demonstrated $B_0$ compensation with fully connected layers. Herein, we improved the $B_0$ compensation and included $B_1^+$ inhomogeneity compensation by convolution layers, see Figure 1.

The network was trained with the stochastic-gradient-descent-with-momentum algorithm (23). Learning progress was monitored through the training and validation subsets. Validation tracking took place every 1000 iteration. The network was trained for 229 epochs (approximately 36 hours), and then manually stopped when the validation RMSE (for DL output against DL target) seemed to converge. The mini-batch size and initial learn rate were 1024 and 0.006, respectively. DL input/target pairs were shuffled at every epoch. The training settings selected differently from MATLAB's default settings, were calibrated by monitoring and minimizing the validation RMSE in trial runs.

<Figure 1>

## 2.4 Experiments

MRI was performed on a 7 T scanner (Magnetom 7T, Siemens Healthineers, Erlangen, Germany), equipped with an 8-channel transmit array (1 kW peak power per channel) and a whole-body gradient system, which can achieve a maximum amplitude of 40 mT/m and slew rate of 200 T/m/s. The custom-built 8-channel transceiver head coil used in this study (24) was driven with circularly polarized (CP) phase setting mimicking a single-channel system. The experimental protocol elaborated below consisted of mapping the $B_0$ and the resulting single-channel $B_1^+$ fields. These were then sent to an offline laptop (2.9 GHz Intel Core i7, 16 GB RAM) for DL-prediction and TM-computation. 2D RF pulses for experiments were then sent back to the scanner console machine for use in the 2D RF pulse sequence including the spiral gradient waveform. Lastly, experimental images were corrected for $B_1^-$ (receive) inhomogeneity.

**2.4.1 Field Mapping**

For absolute $B_1^+$ mapping, a 3D GRE echo scan was acquired in CP mode for two interleaved TRs to compute absolute $B_1^+$ maps according to the AFI approach (25). Parameters were: nominal FA = 90°, TE/TR$_1$/TR$_2$ = 2.64/20/120 ms, FOV$_{read}$ = 250 mm, base-resolution = 64, slice-thickness = 4 mm, slices = 48, phase-resolution = 100% and a bandwidth = 490 Hz/Px. The obtained flip angle was computed following (25). In addition, a relative 3D $B_1^-$ estimate was computed from a second 3D GRE sequence following (26,27), assuming that the sum of magnitudes of all transmit channels is equal to the sum of magnitudes of all receive channels. Parameters were: nominal FA = 5°, TE/TR = 2.37/500 ms, FOV$_{read}$ = 250 mm, base-resolution = 64, slice-thickness = 4 mm, slices = 48, phase-resolution = 100% and bandwidth = 501 Hz/Px.

For $B_0$ mapping, data was acquired with a third 3D GRE sequence with two echo times. Parameters were: nominal FA = 30°, TE$_1$/TE$_2$/TR = 3.06/4.08/500 ms, FOV$_{read}$ = 250 mm, base-resolution = 64, slice-thickness = 4 mm, slices = 48, phase-resolution = 100% and a bandwidth = 596 Hz/Px. The $B_0$ map was computed offline based on the obtained phase differences.

**2.4.2 2D RF experiments**

For 2D RF experiments, specified further below, we inserted the pulse including the spiral gradient waveform into a high-resolution, Cartesian 3D GRE sequence. The following parameters were used: nominal FA = 30°, TE/TR = 1.79/11 ms, FOV$_{read}$ = 250 mm, base-resolution = 256, slice-thickness = 2.08 mm, slices = 96, phase-resolution = 70%, slice-resolution = 69% and a bandwidth = 500 Hz/Px to acquire the entire 3D volume in 2 minutes 10 seconds. The reconstructed sum-of-squares data was corrected for receive profile variations using the relative $B_1^-$ estimations.

**2.4.3 Phantom experiments**

A liquid-filled (50% polyvinylpyrrolidon) cylinder with four air tubes was placed in the head coil. To illustrate the influence of $B_0$ inhomogeneity on 2D RF pulse performance, we performed two sets of experiments with different levels of $B_0$ shimming: i) TuneUp shim with a 350-Hz FID linewidth (full-width at half maximum (FWHM)), ii) a custom $B_0$ shim with a 24-Hz FWHM linewidth.

We performed five 2D RF experiments with each shim setting. One with a DL-predicted pulse and four with TM-computed pulses. The TM-computed pulses differed by which actual (measured) field maps, were included in the optimization or replaced by ideal conditions: $\Delta B_0(\vec{r}) = 0$, $B_1^+(\vec{r}) = 1, \forall \vec{r}$. I.e. four combinations. Inherent to DL-predicted pulses, field maps were included.

**2.4.4 In vivo experiments**

The subject was a healthy volunteer (female, age 24) scanned in the supine position according to an approved IRB protocol. The vendor's adjustment routines were used to determine the transmit reference voltage and to improve the $B_0$ homogeneity across the entire head using second-order shimming.

We performed one 2D RF experiment with a DL-predicted pulse. And then one with a TM-computed pulse including $B_1^+$ and $B_0$ maps in the optimization, and finally one excluding $B_1^+$ and $B_0$ maps for ideal conditions in the optimization.

*3. Results*

**3.1 Numerical results**

Figure 2 shows the numerical pulse performance and characteristics in histograms deduced from the test subset (*N* = 100k). On excitation patterns with respect to target profiles, the mean and standard deviation NRMSE's were 6.0±2.4% and 6.8±2.7% for the TM-computed and DL-predicted pulses, respectively.

<Figure 2>

The TM-computed pulses were hard constrained in amplitude (1 kHz, real/imaginary). Thus, none of these pulses exceeded this limit. The TM-computed and DL-predicted pulses had overlapping amplitude trends, see Figure 2 (center). However, 2% of the DL-predicted pulses exceeded the 1-kHz limit (median overshoot, 73 Hz).

Average power was not constrained or regularized, yet 85% of the DL-predicted pulses had an average power below that of the corresponding TM-computed pulse.

The mean prediction time for the test subset was 8.7 ms on the laptop.

**3.2 Phantom experiments**

Phantom experiments are shown in Figure 3 for the TuneUp shim (linewidth 350 Hz). The Supporting Information Figure S1, shows the corresponding dataset with the custom shim setting (linewidth 24 Hz).

<center><Figure 3></center>

**3.3 In vivo experiments**

Figure 4 shows the in vivo experiments and their corresponding simulations. There is a very similar performance of the DL-predicted (Fig. 4, top row) and the TM-computed (Fig. 4, middle row) pulses, when $B_0$ and $B_1^+$ maps are incorporated. When the TM-computed pulse assumes ideal $B_0$ and $B_1^+$ conditions ($\Delta B_0(\vec{r}) = 0$, $B_1^+(\vec{r}) = 1, \forall \vec{r}$) in computation, but experiences actual conditions in a subsequent simulation and naturally in experiments, the excitation quality diminishes (Fig. 4, bottom).

<center><Figure 4></center>

Figure 5 shows the diagonal trace profiles of the three simulations in Figure 4. The diagonal trace was chosen as it happened to possess the widest $B_1^+$ range within the target region. Clearly the actual $B_1^+$ profile scales the response of the TM-computed pulse that assumed an ideal $B_1^+$ profile. This is counteracted by the tailored DL-predicted and TM-computed pulses.

<center><Figure 5></center>

Pulses used in the in vivo experiments are displayed in the Supporting Information Figure S2.

***4. Discussion***

Rapid (< 10 ms) single-channel 2D RF pulse generation by convolutional neural networks featuring $B_0$ and $B_1^+$ inhomogeneity compensation in vivo and in a phantom at 7 T has been presented.

Deep learning the network was based on an input-target supervision library, with target pulses computed with, in this case, an optimal control framework, and inputs being randomly shaped excitation targets, and $B_0$ and $B_1^+$ maps deduced from a photography database, without any MRI heritage, except the images were scaled to appropriate $B_0$ and $B_1^+$ maps by typically observed values from the MRI setting.

As this proof-of-concept serves to demonstrate the networks ability to compensate $B_0$ and $B_1^+$ inhomogeneity, we did not make any effort to optimize the library size. We chose an unusually high number of training/validation/test examples - 500k in total. As shown in (14) this number could potentially be lowered by studying the L-curve, however, building the library was rather effortless because we used a fast pulse computation strategy optimized for parallel execution, computing 28 pulses simultaneously. Partial $B_0$ inhomogeneity compensation was observed in the preliminary studies of (14) using only fully connected layers and up to 60k training examples. Herein, however, we found a leap in performance by including convolution layers.

Judged from the histograms of Figure 2 the DL-predicted and TM-computed pulses had quite overlapping behaviour. Except for an expected slightly lower DL excitation fidelity, which can be further minimized by more training examples, other network structures, and better training strategies (14); we saw a low (median, 7.3%) pulse amplitude overshoot in 2% of the test cases. Suggestions for eliminating such occasional overshoots, which the network is not designed to actively do, were given in (14), e.g.: setting the amplitude hard-limit for the target method at a lower level. Ongoing research seeks to solve this issue rigorously.

The proposed method will be accessible at github.com/madssakre/DeepControl.

With the present optimal control configuration, computations were around half a minute on the laptop used in the experiment session (more than three orders of magnitude longer than the prediction time). Half a minute was still, however, rather benign for our experiments with the volunteer, but computation time can rapidly increase to many minutes by using more spatial control points, multiple RF channels, and longer pulses etc. (7). Or even hours, when including exact gradients and local SAR constraints (7), or multiple shim channels (28). In these cases, although the neural network complexity would increase too, we would still expect the prediction time to be several orders of magnitude faster than a conventional approach. We are currently working on extending the framework to parallel transmit pulses with acceleration for shorter pulse duration.

**5. Conclusion**

Our experiments (in vivo at 7 T) and simulations show that the proposed, trained convolutional neural network predicts 2D RF pulses, which do actively compensate $B_0$ inhomogeneity (sharpness of the excitation profile) and $B_1^+$ insensitivity (flip angle homogenization), similarly to a tailored pulse computed by a conventional method. The training library is essentially free to build and improve. The pulse prediction time currently less than 10 ms is essential for clinical practice.

**6. Acknowledgement**

We thank Villum Fonden, Eva og Henry Fraenkels Mindefond, Harboefonden, and Kong Christian Den Tiendes Fond. We also thank MSc Birk Skyum for helping develop the network.

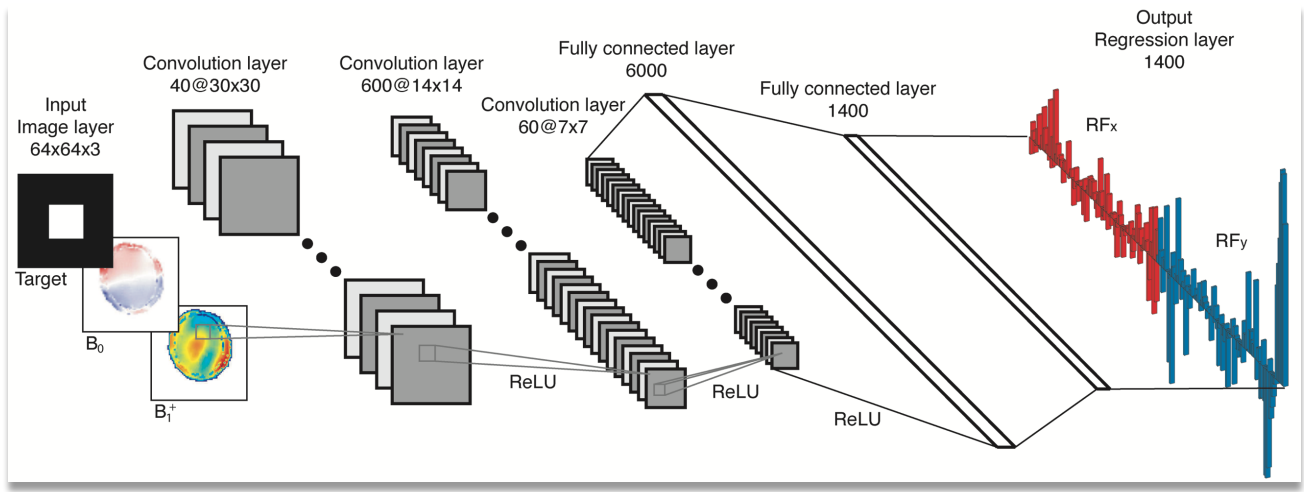

**Figure 1:** The proposed convolutional neural network for 2D RF pulse prediction with $B_0$ and $B_1^+$ inhomogeneity compensation. Left side constitutes the TM/DL input as a 64x64x3 matrix (see text). Three convolution layers with rectified linear unit (ReLU) functions are followed by two fully connected layers, and finally a regression layer constitutes the DL output. The 1400-length array of numbers is split in two halves constituting the real and imaginary RF pulse waveforms, making a complex RF pulse of 7-ms length. The RF pulse is played together with a fixed gradient waveform (spiral trajectory). For training purposes described further in (14), the DL inputs and targets were scaled to have values strictly in the range -1 to 1, i.e., the $B_0$ maps and 2D RF pulses were normalized to 600 Hz and 1 kHz, respectively.

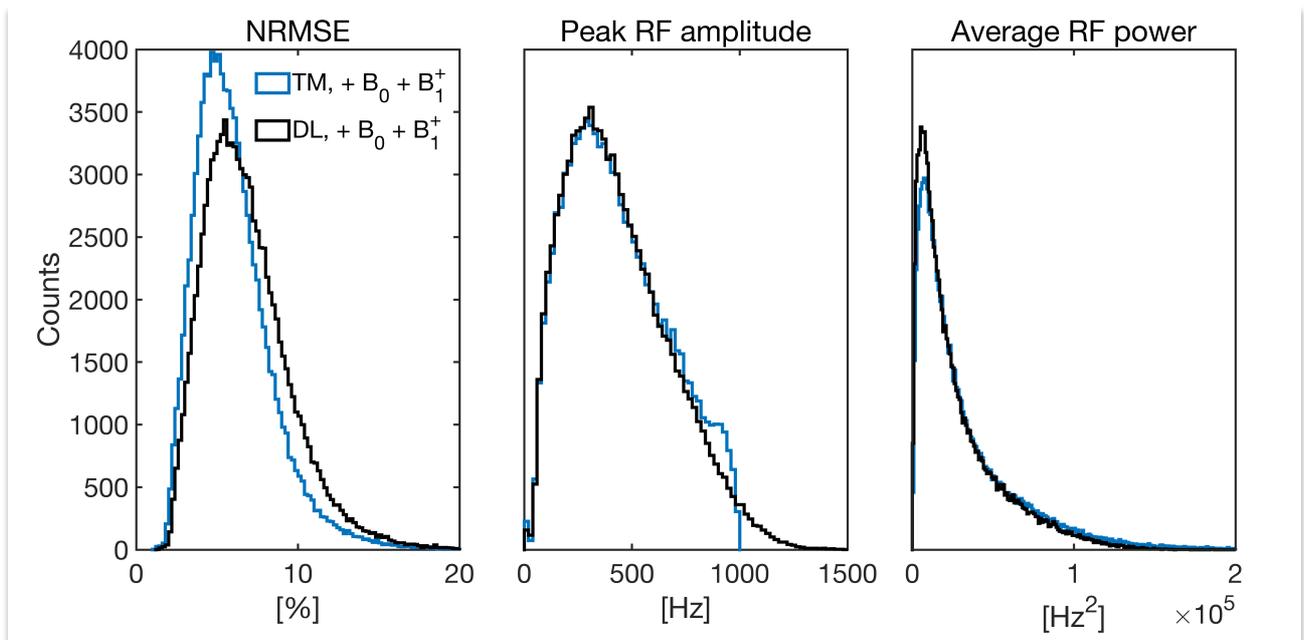

**Figure 2:** Histograms of numerical results (*N* = 100k). Left, NRMSE of the actual excitation with respect to the target excitation pattern. Center, peak RF amplitude, showing the hard limit at 1 kHz for the TM-computed pulse. The tail beyond 1 kHz for the DL-predicted pulses account for 2 % of the cases. Right, average RF power, which for DL-predicted pulses are below the corresponding TM-computed pulses in 85 % of the cases.

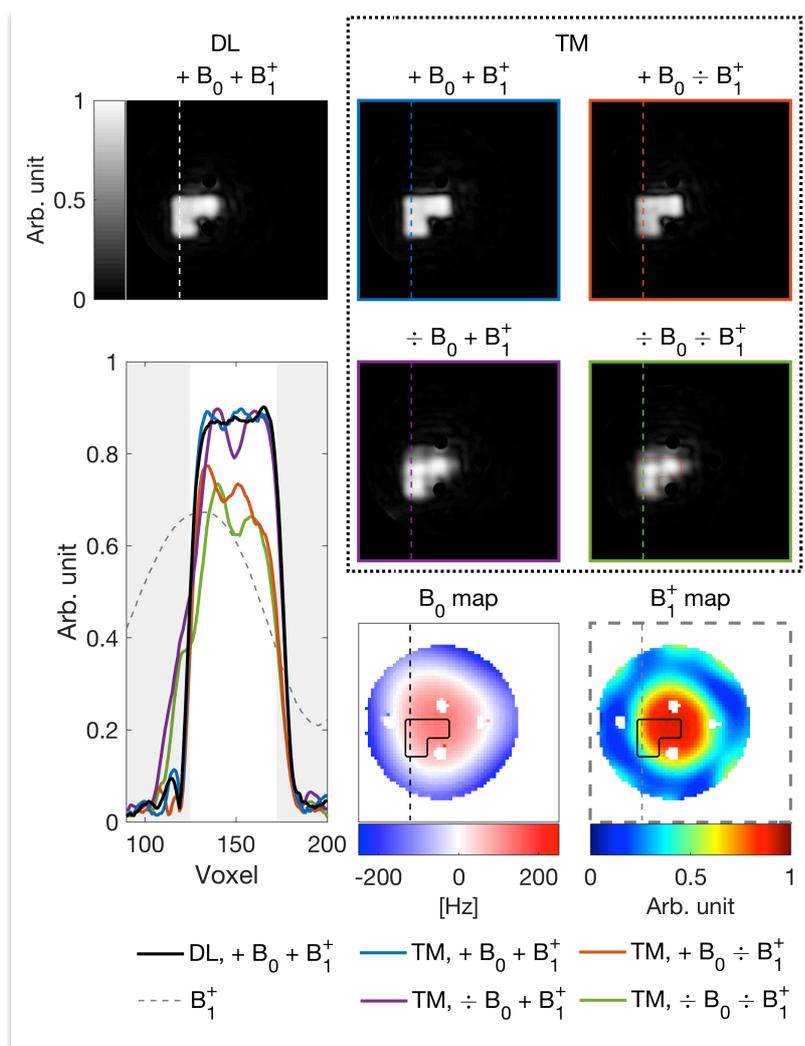

**Figure 3:** Phantom experimental data, with a deliberately poor TuneUp shim having an FID spectral linewidth of 350 Hz (FWHM). Profiles in the lower left plot correspond to the vertical, dashed lines in the images. The shaded regions correspond to the outside of the target. The nomenclature of e.g. + $B_0$ ÷ $B_1^+$ refers to the situation, where the measured $B_0$ map was included in the pulse optimization, but the measured $B_1^+$ map was excluded and replaced by ideal $B_1^+$ conditions. DL-predicted pulses inherently have both maps included, but the TM-computed pulses were tested with the all four map combinations. The frame coloring around the TM-computed pulse images are also used in the profiles plot, for convenience. It is clearly seen that exclusion of the $B_0$ map in the optimization results primarily in a blunt profile, whereas exclusion of the $B_1^+$ map results primarily in a lower flip angle scaled according to the $B_1^+$ profile. The DL-predicted pulse largely follows the profile of the TM-computed pulse, where both maps were present in the optimization. The four holes in the maps correspond to the air tubes within the phantom. Supporting Information Figure S1, shows the data for the experiment with an improved shim, i.e., with the FID spectral linewidth of 24 Hz (FWHM).

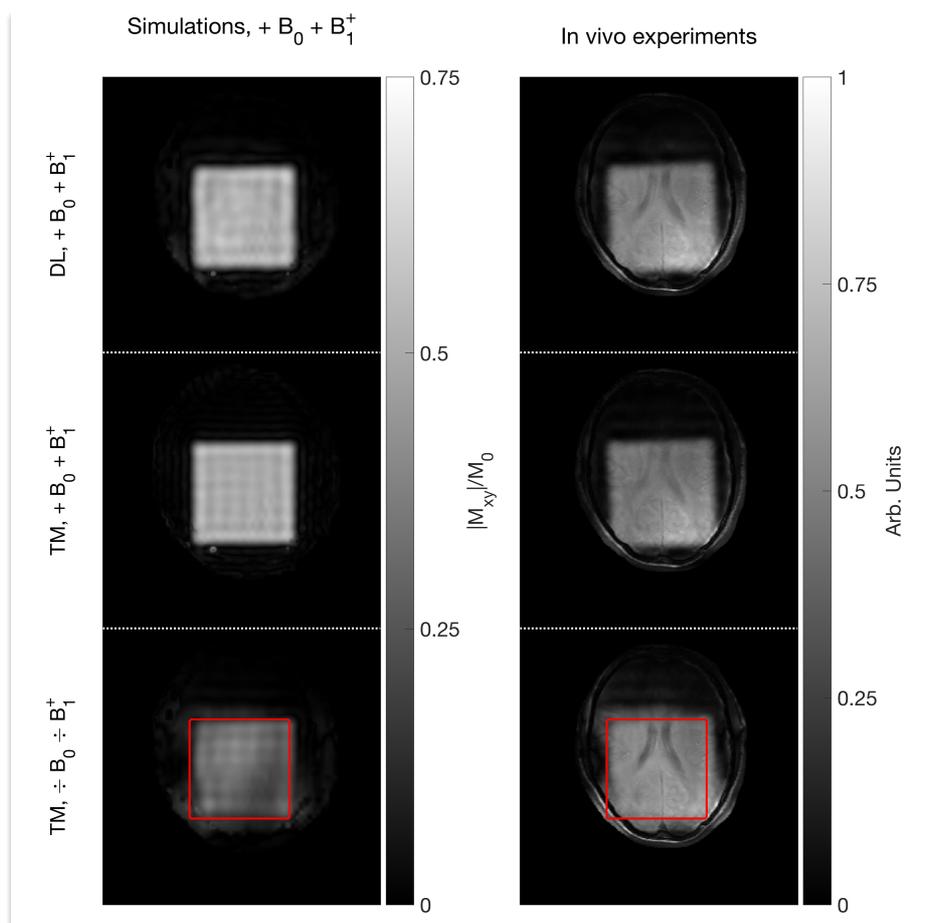

**Figure 4:** Experimental results (right column) with the corresponding simulations (left column). The top row shows the results from the DL-predicted pulse. The middle row is from the TM-computed pulse where the measured $B_0$ and $B_1^+$ maps were included in the computation and simulation. The bottom row shows the results from the TM-computed pulse, where the measured $B_0$ and $B_1^+$ maps were not included in the computation, but included in the simulation (bottom left). The $B_0$ and $B_1^+$ maps are shown in Figure 5. The pulse sequence did not include a fat saturation module, why the skull lights up (29,30). The red boxes correspond to the target region.

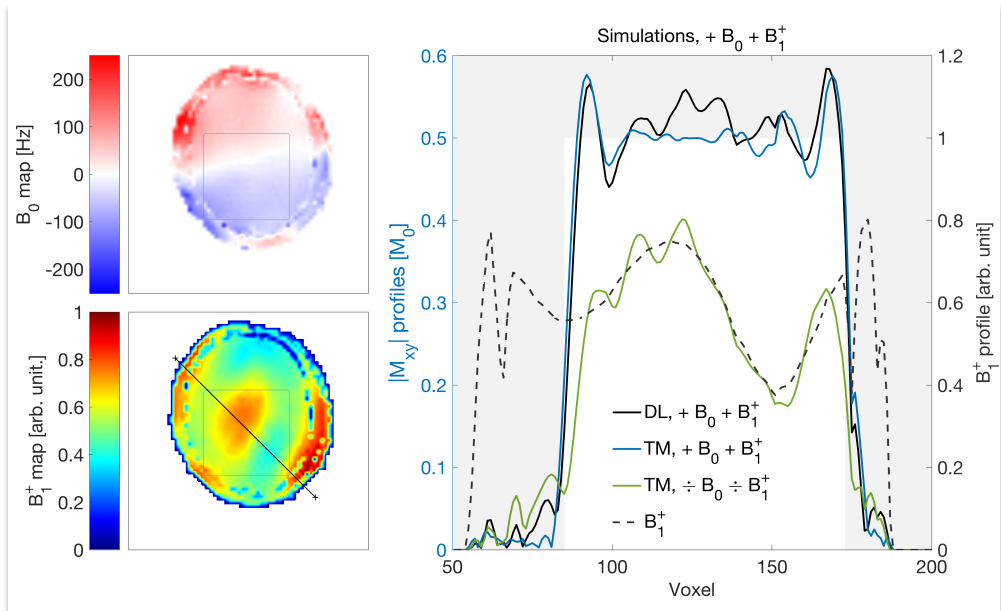

**Figure 5:** $B_0$ and $B_1^+$ maps (left column) and magnetization profiles (right plot). The magnetization profiles stem from the plots of Figure 4 (left column, simulations) of the diagonal trace line depicted in the $B_1^+$ map. Also shown is the corresponding $B_1^+$ map profile. The overlaid box in the maps and the shaded area in the profile plot resemble the target region (black/white image in Figure 1). The choice of trace line was found by inspecting, which trace line (checked for 512 traces rotated in total 180° about the centre), that had the highest dynamic range in $B_1^+$ values resulting by coincidence in the diagonal trace. The highest dynamical range of $B_0$ values was observed on a trace rotated approximately 3° off the given trace.

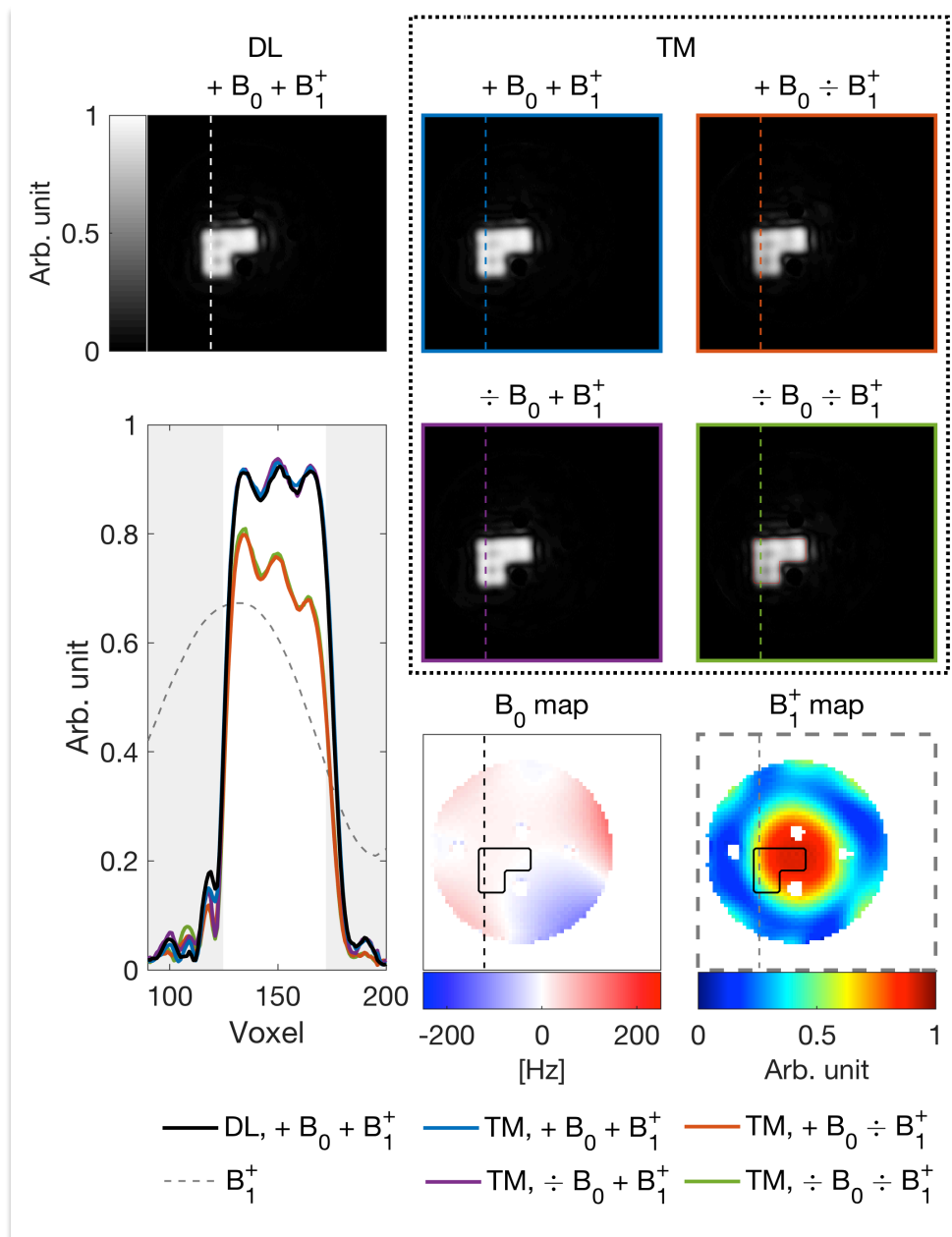

**Figure S1:** Phantom experimental data, with a better cumstomized shim having an FID spectral linewidth of 24 Hz (FWHM). Profiles in the lower left plot correspond to the vertical, dashed lines in the images. The shaded regions correspond to the outside of the target. The nomenclature of e.g. + $B_0 \div B_1^+$ refers to the situation, where the measured $B_0$ map was included in the pulse optimization, but the measured $B_1^+$ map was excluded and replaced by ideal $B_1^+$ conditions. DL-predicted pulses inherently have both maps included, but the TM-computed pulses were tested with the all four map combinations. The frame coloring around the TM-computed pulse images are also used in the profiles plot, for convenience. It is clearly seen, compared to Figure 3, that exclusion of the $B_0$ map

in the optimization has a lower effect (with a better shim), whereas exclusion of the $B_1^+$ map results again primarily in a lower flip angle scaled according to the $B_1^+$ profile. The DL-predicted pulse largely follows the profile of the TM-computed pulses, where both maps or at least the $B_0$ map were present in the optimization. The four holes in the maps correspond to the air tubes within the phantom.

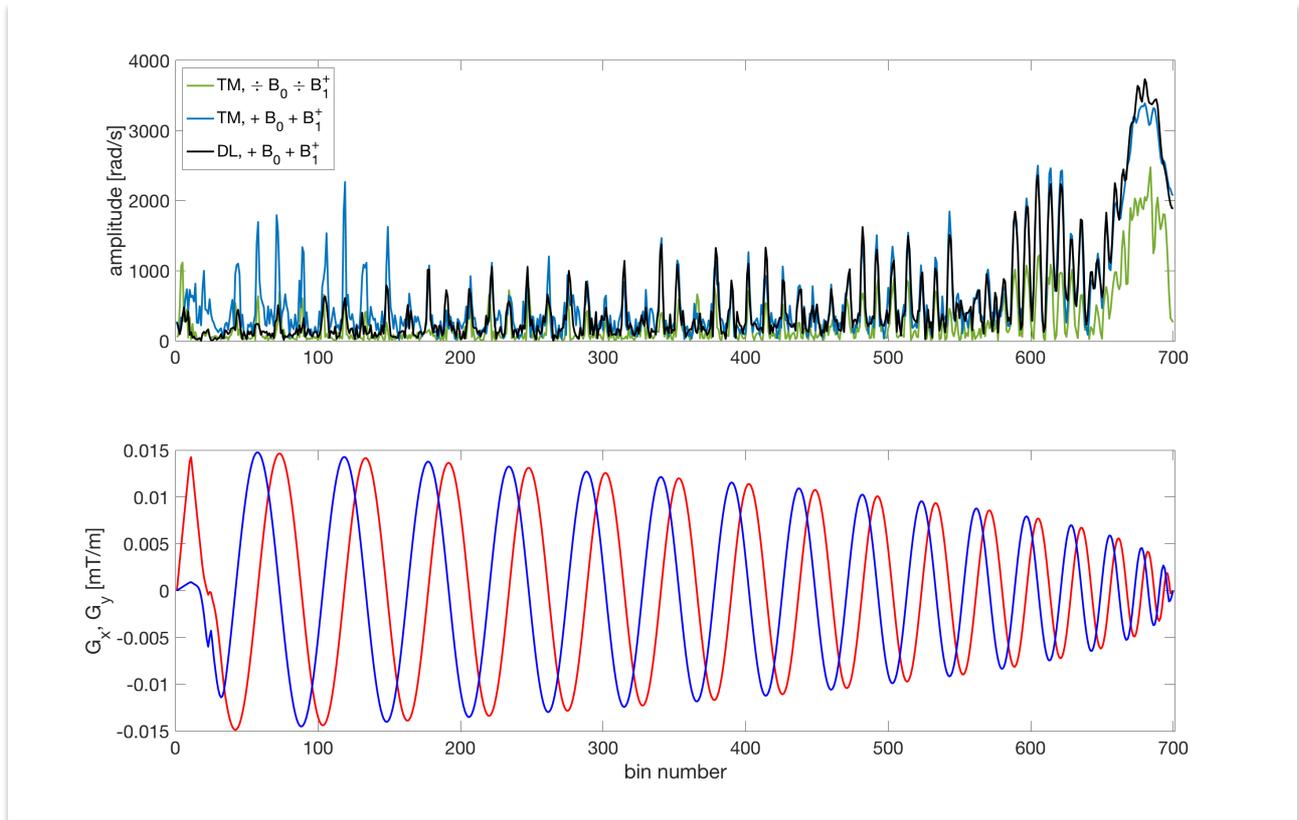

**Figure S2:** RF and gradient pulse waveforms from the in vivo experiments. All pulses were scaled internally on the Siemens system in the in vivo experiments to yield a nominal flip angle of 30 degrees.